# Microfluidic pumping using artificial magnetic cilia

Srinivas Hanasoge - srinivasgh@gatech.edu
Peter J. Hesketh - peter.hesketh@me.gatech.edu
Alexander Alexeev - alexander.alexeev@me.gatech.edu

George W. Woodruff school of mechanical engineering,
Georgia Institute of Technology, Atlanta, GA, USA, 30332

Corresponding author: peter.hesketh@me.gatech.edu

**Abstract**

One of the vital functions of naturally occurring cilia is fluid transport. Biological cilia use spatially asymmetric strokes to generate a net fluid flow that can be utilized for feeding, swimming, and other functions. Biomimetic synthetic cilia with similar asymmetric beating can be useful for fluid manipulations in lab-on-chip devices. In this paper, we demonstrate the microfluidic pumping by magnetically-actuated synthetic cilia arranged in multi-row arrays. We use a microchannel loop to visualize flow created by the ciliary array and to examine pumping for a range of cilia and microchannel parameters. We show that magnetic cilia can achieve flow rates of up to $11 \mu l/min$ with the pressure drop of $\sim 1 Pa$. Such magnetic ciliary array can be useful in microfluidic applications requiring rapid and controlled fluid transport.

**Keywords –** Biomimetic artificial cilia, microfluidic pumping, active microfluidic transport.

1. Introduction

Achieving fluid transport at the microscale is particularly difficult owing to the lack of inertial effects[1,2]. It means that any reciprocal motion displaces the fluid back and forth without producing any net flow[3]. Microorganisms typically operate in these regimes of negligible inertial effects. To perform vital biophysical functions requiring fluid agitation and motion, microorganisms have evolved to use cilia and flagella with complex beating patterns yielding a net fluid transport[4–7]. Ciliary bands are used to produce feeding currents that draws the food particles towards the mouth of the organism for feeding[8,9]. Cilia facilitate organism swimming[9,10]. Further, the flow produced by motile cilia in an embryo is shown to directly affect the developing organism left-right asymmetry[11]. The cilia beating can be planar or three-dimensional[12]. Researchers have studied flow patterns generated by natural cilia[8–11,13–15] and have shown that ciliary carpets[14,15] are able to produce flow speeds of up $1 mm/s$. These biological cilia inspire researchers to develop synthetic analogs of cilia capable of performing complex biomimetic functions that could prove to be useful for lab-on-chip microfluidic devices, in vitro and in vivo artificial organs, and drug delivery applications[16–23].

Different fabrication methods have been demonstrated to create artificial cilia capable of microfluidic pumping[24–26]. Researchers have used continuous roll up approach[24] to fabricate synthetic cilia and have demonstrated transitional flow speeds up to $120 \mu m/s$ in such systems. The self-pumping frequency – a metric used to assess the effectiveness of the pump based on its size[27] (ratio of the flowrate to package size $f_s = Q/S_{package}$) is estimated to be $0.2 \min^{-1}$ for this pump. Wang et al[25] have used self-assembled magnetic bead cilia to demonstrate transitional flow speeds of $\sim 240 \mu m/s$. Toonder et al.[28] fabricated micro beams from bilayer films of polyimide and chromium. Their cilia showed substantial mixing and pumping abilities when actuated electrostatically, and demonstrated a translational fluid flow rate of over $500 \mu m/s$. Fahrni et al.[26] made ferromagnetic PDMS flaps using photolithography techniques and actuated them externally using an electromagnetic setup with a frequency up to $50 Hz$. Rotational as well as translational flow was created with instantaneous fluid velocities of up to $\sim 500 \mu m/s$. Hussong et al.[29] have shown the use of polymeric cilia capable of producing net transitional velocities of up to $\sim 130 \mu m/s$. In general, ciliary pumps have shown to produce a relatively high flowrate, whereas the pressure drop in these systems is relatively low ($\sim 1Pa$). Among different fabrication and actuation techniques demonstrated for artificial cilia, the use of magnetic cilia is highly promising due to the ease of actuation by manipulating an external magnetic field. Further, magnetic actuation does not interfere with the biological samples making such cilia especially useful in various biomedical applications.

Researchers have studied the mechanism of artificial cilia beating and fluid transport through computer simulations[21,30–32]. Khadri et al.[33] have simulated the pumping characteristics of magnetic cilia in a microchannel. They show results with cilia capable of producing flow rates of $\sim 18 \mu l/min$ and $3mm$ of water pressure. The net flow generated by magnetic cilia and its relation to the beating pattern and Reynolds number was analyzed[34]. Researchers have also examined the effects of multiple cilia interacting in an array and discussed the effects of metachronal waves[35–37]. The use of cilia for micro-particle capture[20,23,38,39], particle transport[40], and flow control[21] has been examined using computational modeling.

A magnetic filament actuated by a rotating magnetic field is shown to produce spatially asymmetric motion[41] in a two-dimensional plane of oscillation. Furthermore, this motion is time-irreversible, in that, reversing the direction of magnet rotation results in a different beating pattern. In this work, we probe the effect of these different beating pattern, obtained by reversing the rotation of the magnetic field, on fluid pumping. We examine the two-dimensional motion of artificial cilia and study the net pumping generated by an array of synchronously actuated cilia. To quantify the pumping, we incorporate an array of cilia inside a closed loop micro-channel. Thus, actuated cilia pump and circulate fluid through the channel loop. The pumping rate is measured by visualizing the flow. The combined effect of multiple beating cilia generates rapid microfluidic pumping with rates that depend on parameters including the direction of magnet rotation, the cilia properties, dimensions of ciliary array and the microchannel dimensions. Our cilia arrays can generate transitional flow speeds of up to $1.4 mm/s$ corresponding to a flowrate of $11 \mu l/min$ in closed loop channels. To the best of our knowledge, these results are the highest reported flowrates for such ciliary pumps. Our magnetic cilia can find applications in biomedical assays. Closed loop pumping systems are useful in cell culture applications, where a fluid sample needs to be circulated continuously. They can potentially replace conventional peristaltic pumps, which are bulky and expensive.

## 2. Methods

*2.1 Fabrication*

We employ surface micromachining techniques involving a simple two-mask lithographic process. The first lithography step results in wells in the photoresist. These features are then deposited with a sacrificial layer of copper, followed by a layer of nickel-iron. Lift-off is done to remove the photoresist by dissolving it in acetone, which leaves the cilia features on the surface. The next step is to deposit an anchor for holding the cilia on the substrate. A second lithographic step is performed to obtain these features which are deposited with titanium. This anchor layer sticks to the glass and anchors the NiFe cilia to the substrate. The cilia are then released by removing the sacrificial copper layer by dissolving it in 5% ammonium hydroxide, which selectively etches the copper. This leaves a free standing magnetic NiFe film that can be actuated by an external magnetic field. This process is simple, involves standard procedures, and is highly reproducible. Moreover, the fabricated devices can be stored for extended periods of time, and the sacrificial layer can be removed during the time of use. The dimensions of the cilia are readily modified by changing the lithographic mask and NiFe film deposition conditions. Further details of the fabrication process can be found elsewhere[41,42].

*2.2 Setup and experimental protocol of pumping experiments*

To study the fluid pumped by the cilia, we incorporated an array of cilia inside a microchannel loop. The dimensions of the cilia are length $L = 200 \mu m$, width $W = 20 \mu m$, and thickness $T = 60 nm$, whereas the spacing between neighboring cilia within a row is $20 \mu m$. A PDMS microchannel loop (width of $1mm$ and depth of $300 \mu m$) is fabricated using standard soft lithography[43] and two inlet/outlet holes are punched for adding and removing the fluid. The channel is then placed on the glass substrate decorated with a micro-machined ciliary array, and gently pressed to form a reversible bond.

After bonding the PDMS on glass surface, pure ethanol is introduced to wet the inner walls of the channel. DI water is then introduced in the already wet channel to displace ethanol. This is followed by introducing 5% ammonium hydroxide to selectively etch the copper sacrificial layer and release the cilia. The last step is to displace the ammonia water with DI water containing fluorescent beads ($3 \mu m$, Fluoro-Max, Thermo-Fisher scientific) for flow visualization. Lastly, the inlet/outlet reservoirs are sealed by placing a glass coverslip on top of the PDMS and gently pressing. This process results in a leak free channel loop with cilia on the bottom channel wall. Videos of fluorescent beads are recorded under a microscope to evaluate the fluid flow velocity. A schematic of the experimental setup is shown in Fig. 1.

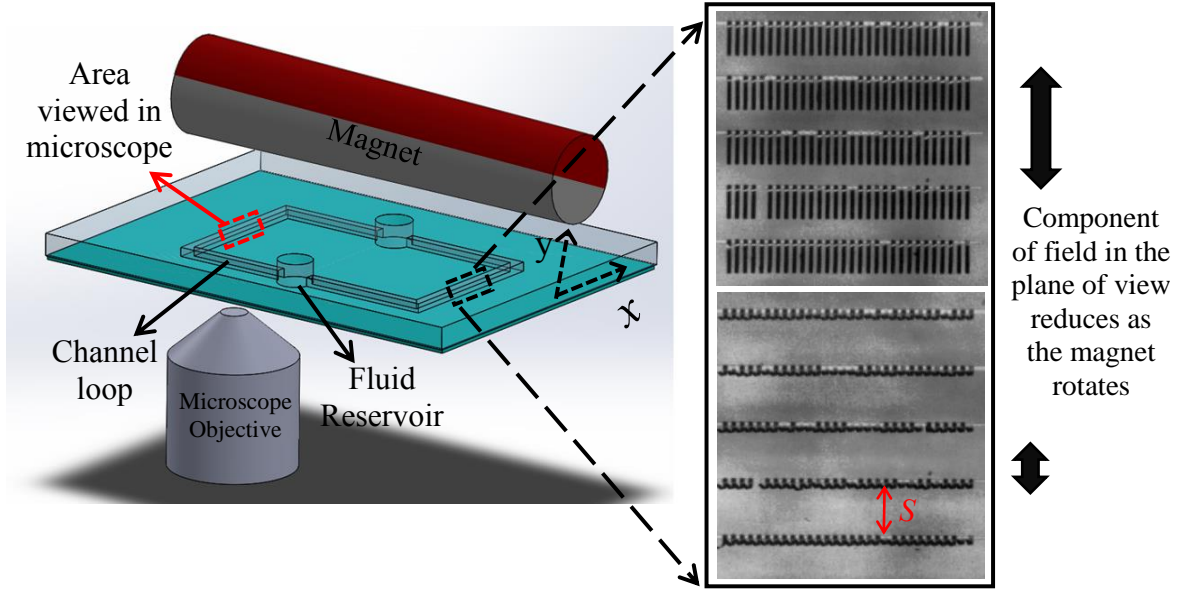

**Figure 1**: Schematic of experimental setup showing a microchannel loop through which the fluid pumped by cilia is circulated. A large diametrically magnetized permanent magnet is used to actuate the cilia. Inset shows actuated ciliary array as seen from the top for two different positions of the magnetic field.

A long permanent magnet, magnetized through its diameter and rotated along the long axis is used to actuate the ciliary array. The magnet is held in the position using a fixture that can be precisely manipulated in the $x$ and $y$ directions such that the magnet axis aligns with the center of the array. The diameter of the permanent magnet is $D = 12.5mm$, which is significantly larger than the size of an individual cilium with is about $100 \mu m$. This ensures that each cilium is exposed to a nearly uniform magnetic field. Figure 1 shows the top view of a ciliary array at two positions of the beating cycle. All cilia show the same in-phase motion due to their synchronous actuation. See video in ESI.

In our experiments, we study the pumping by varying the frequency, direction of rotation, number of cilia rows, spacing between rows, and the channel height. The pumping produced by ciliary array is characterized by measuring the centerline flow velocity on the opposite side of the microfluidic channel loop (see section A in Fig. 1). To measure this velocity, we recorded videos of fluorescent particles flowing in the microchannel. Care was taken to ensure that the particles near the center of the microchannel were imaged. ImageJ is used to track particles by manually selecting the start and end positions. The centerline velocity is computed by measuring the distance travelled by the particles in a given time.

*2.3 Imaging of cilium beating pattern*

To study cilium beating kinematics, the beating pattern of a cilium is imaged in the plane of its oscillation. To this end, we place the substrate such that the cilium oscillates in the plane of view of the microscope as described previously[41]. A permanent magnet is rotated with its axis along the vertical direction producing a magnetic field that rotates in the horizontal plane. A high-speed camera is used to image the trajectory of the cilium. A cilium of length $500 \mu m$, width $10 \mu m$ and thickness $70 nm$ is used for imaging the side view.

*2.4 Flow simulation*

We use two-dimensional computer simulations to understand the kinematics of a single cilium. The fluid-structure interaction simulations is performed using COMSOL with an arbitrary Lagrangian-Eulerian (ALE) method[44]. To model magnetic actuation, we impose a uniform magnetic field giving rise to a distributed moment acting on the elastic cilium. We locally apply magnetic moment along the cilium with the magnitude that is proportional to $\sin(2\theta)$, where $\theta$ is the angle between magnetic field **B** and the local axis of the cilium[17,33,41,45,46]. The magnitude of the moment is set to match cilium deflection observed in the experiment. The fluid far from the cilium is assumed to be stationary. Results from the simulations are compared against the beating kinematics of a single isolated cilium.

*2.5 Non-dimensional parameters*

We introduce non-dimensional parameters to characterize the operation of the cilia. The Reynolds number of the beating cilia is $Re = \rho L W f / \mu$ where, $\rho$ is the density, $L$ is the length of the cilium, $W$ is the width, $f$ is the frequency of operation, and $\mu$ is the viscosity of water. We estimate $Re$ to be in the range between 0.09 and 0.6 for all our experiments. For these relatively low values of $Re$, the cilia operate in a flow regime that is dominated by viscous forces[2].

It has been shown that for $Re \ll 1$ the kinematics of a cilium depends entirely on a magnetic number $Mn$ and a sperm number $Sp$.[34,41,47] The sperm number $Sp = L(\omega\xi/EI)^{0.25}$ characterizes the ratio of viscous to elastic forces acting on the cilia. Here $\omega = 2\pi f$ is the angular velocity, $\xi = 4\pi\mu$ is the lateral drag coefficient of the cilium, and $EI$ is the cilium bending rigidity. In our experiments we varied $Sp$ by varying the oscillation frequency.

The magnetic number $Mn = (B^2 L^2 WT/\mu_0 EI)^{0.5}$ represents the ratio of magnetic and elastic forces acting on the cilia[30,41,45,47]. Here, $B = |\mathbf{B}|$ is the magnitude of the magnetic flux density and $\mu_0 = 4\pi \times 10^{-7}$ is the permittivity of free space. In the experiments reported here, the magnetic number is kept constant and equal to $Mn \approx 3.1$.

To quantify the pumping by arrays of magnetic cilia, we define a dimensionless parameter $P_f = U_0 Sp^4/\omega L$ that represents pumping generated by the ciliary array per unit time[21]. Here, $U_0$ is the centerline velocity measured in segment $A$ of the channel (Fig. 1).

### 3. Results and discussion

The soft magnetic cilia get magnetized in the presence of an external magnetic field and tend to align along the direction of the field. We make use of this to actuate the cilia by subjecting it to a magnetic field rotated in the *x-y* plane as shown in Fig. 1. The magnetic field is rotated either in the clockwise (CW) or in the counter clockwise (CCW) direction. The change of the direction of magnet rotation results in significantly different beating patterns as shown Fig. 2 and in Fig. 3 that present the side view images of cilia collected during CCW and CW magnet rotation, respectively. In the following sections, we first examine the cilia beating due to CCW and CW magnet rotation, and then probe how the cilia beating translates in to fluid pumping in a microfluidic channel.

*3.1. Counter-clockwise rotation of magnetic field:*

Magnetic cilia follow spatially asymmetric motion when exposed to a CCW rotating magnetic field[41]. An example of such a motion is shown in Fig. 2a that presents a series of overlapped images of a cilium actuated by a CCW rotating magnet. We separate the spatially asymmetric beating pattern of cilia into the forward and backward strokes. The forward stroke is defined as the motion of cilium from position *a* to *d*. The backward stroke is defined as motion of cilium as it returns from position *d* to *a*. Note that position *a* does not coincide with the position the cilium assumes without the influence of the magnetic force which is nearly horizontal, but rather it is defined as the rightmost position of the cilium through a beating cycle. The cilium tip follows a closed trajectory shown in blue during the forward stroke and in red during the recovery stroke. Differences in cilium bending pattern during the forward and recovery strokes leads to a high degree of spatial asymmetry of cilia motion.

Computer simulations performed for the CCW rotation of magnetic field predict a beating pattern that closely agrees with the experimental results (Fig. 2b). We use computer simulations to closely examine the motion of a cilium due to the rotating magnetic field. We start at position *a* where the cilium is at its right most position corresponding to the beginning of the forward stroke. At this position, most of the cilium is aligned with the external magnetic field, except for a small portion near the cilium base. The local magnetic moment is negligible along a large section of the cilium (Fig. 2c). Remember, the local magnetic moment is proportional to $\sin(2\theta)$, where $\theta$ is the local angle between cilium and magnetic field. As the magnet rotates in the CCW direction, $\theta$ increases, increasing the magnitude of the magnetic moment. As a result of this CCW rotation, the cilium deforms and bends in CCW direction.

The cilium proceeds through positions *b-c-d* in Fig. 2b following the rotation of the external magnetic field. The bending results in the local angle $\theta$ to exceed $90^0$ at certain parts of the cilium. Thus, the base of the cilium in position *c* makes an angle $\theta > 90^0$, whereas the cilium tip that more easily deflects to follow the magnetic field makes an angle $\theta < 90^0$. This results in a situation in which the direction of the local magnetic moment near the base is opposite to that at the cilium tip (Fig. 2c). This change in the moment direction along cilium further enhances the deformation. The bending of cilium from *a* to *d* increases the elastic energy stored by cilium, which reaches a maximum at position *d*, where the cilium reaches the maximum bending. Further rotation of the magnetic field reduces the applied moment and the cilium recovers to the initial position *a* via position *e*. During the recovery stroke, the magnetization of the cilium flips direction and so does the applied moment, which now acts to pull the cilium towards –**B** (see Fig.1c for position *e*). The stored elastic energy is released driving the cilium to quickly snap back to its initial position *a*. Due to the fast cilium motion during recovery stroke, the viscous forces are significantly higher than that during forward stroke where the velocity is defined by the rate of magnetic field rotation. Furthermore, the magnetic moment flips direction during recovery and thus aids in the return of cilium to its initial position. As a result, the time scale for recovery stroke is much smaller than that of the forward stroke, which in turn leads to an asymmetric beating. A more comprehensive discussion on the kinematics of cilium motion in CCW mode through experiments and simulations can be found elsewhere[30,32,41].

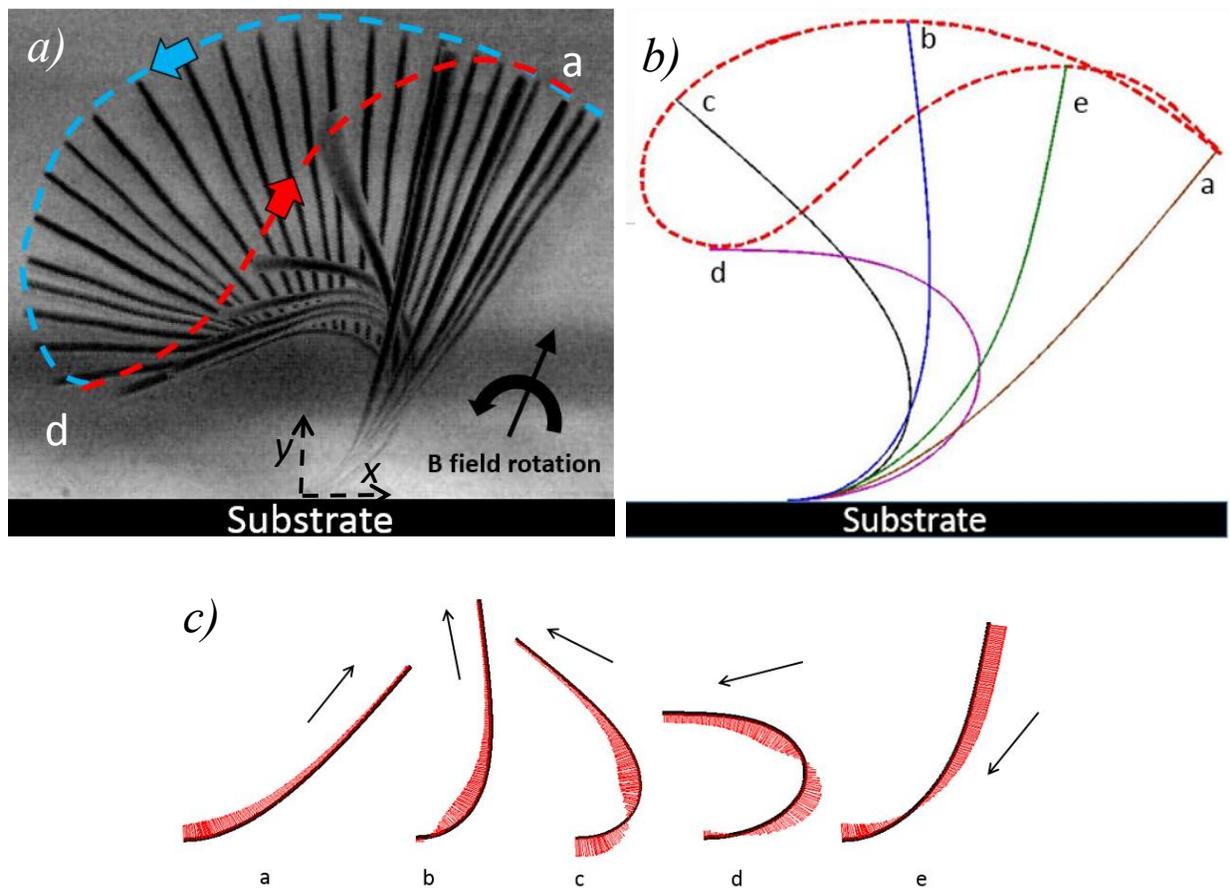

**Figure 2**: Cilium beating due to counter-clockwise (CCW) rotation of magnetic field. (**a**) Overlay of experimental images showing cilium profiles for different orientations of magnetic field. Experimental parameters are $L = 500\mu m$, $W = 10\mu m$, $T = 70nm$, $\omega = 0.3Hz$. (**b**) Computer simulation of cilium motion induced by CCW rotating magnetic field. (**c**) Magnitude of local bending moment acting on the cilium at various orientations of magnetic field. The arrows indicate the field direction.

*3.2. Clockwise rotation of magnetic field*

Figure 3a shows a cilium that is subjected to a magnetic field rotating in the CW direction. This cilium has the same properties as the one shown in Fig. 2a. We find that cilium beating due to CW magnet rotation differs significantly from that induced by the CCW rotation of the magnet (cf. Fig. 2a and Fig. 3a). In particular, the limiting positions *a* and *d* between which the cilium oscillates are shifted to the right when the magnet rotation is changed from CCW to CW (i.e., the cilia tend to deform in the direction of the magnet rotation). Furthermore, the area enclosed by cilium tip is significantly smaller for the CW rotating magnet than for the CCW rotation. This indicates that the CCW magnet rotation is beneficial for enhancing beating asymmetry.

Cilium motion obtained in the experiments (Fig. 3a) is well captured by the computer simulations (Fig. 3b). We define the CW forward stroke to begin at position *a*. At this position most of the cilium is aligned with the field direction, except near the base. As the magnetic field rotates in the CW direction, the magnetic moment decreases cilium bending as shown in Fig. 3c. The cilium follows the rotating field until position *d* when it touches the substrate. The magnetic moment at this position is negligible since the cilium is fully aligned with the field. The cilium remains at this position *d* prevented by the substrate to move as the magnetic field continues to rotate CW. Indeed, until the angle $\theta$ between the cilium and field is less than $90^0$, the moments act in the clockwise direction, and force the cilium to remain in position *d*.

When the magnetic field is oriented vertically, $\theta$ is equal to $90^0$. Further rotation of the field flips the magnetization in the cilium and leads to a situation where the angle between cilium and $-\mathbf{B}$ reduces. This is indicated by *d'* in Fig. 3c. The cilium does not move and is aligned along the substrate between positions *d* and *d'*. The magnetic moments at *d'* flips direction and pulls the cilium upwards and away from the substrate. As the cilium bends to align with the magnetic field, it goes thorough an intermediate position *e* and completes the recovery stroke back at the nearly vertical position *a*. Thus, during the recovery stroke the cilium moves in the direction opposite to the rotation of the field.

Similar to CCW field rotation scenario, cilium speed during forward stroke is set by the rate of the CW magnetic field rotation. During the backward stroke, cilium motion is defined by the balance between magnetic force and viscous forces. (Recall that for CCW rotation, the recovery stroke is a result of balance between the elastic and viscous forces) The difference in acting forces between the forward and recovery strokes gives rise to the asymmetry of the cilium motion in CW rotating magnetic field. Note that at the end of the forward stroke the cilium pauses while the magnetic field continues to rotate clockwise until the magnetization in the cilium changes direction. No such pause exists for cilia driven by the CCW rotating magnetic field.

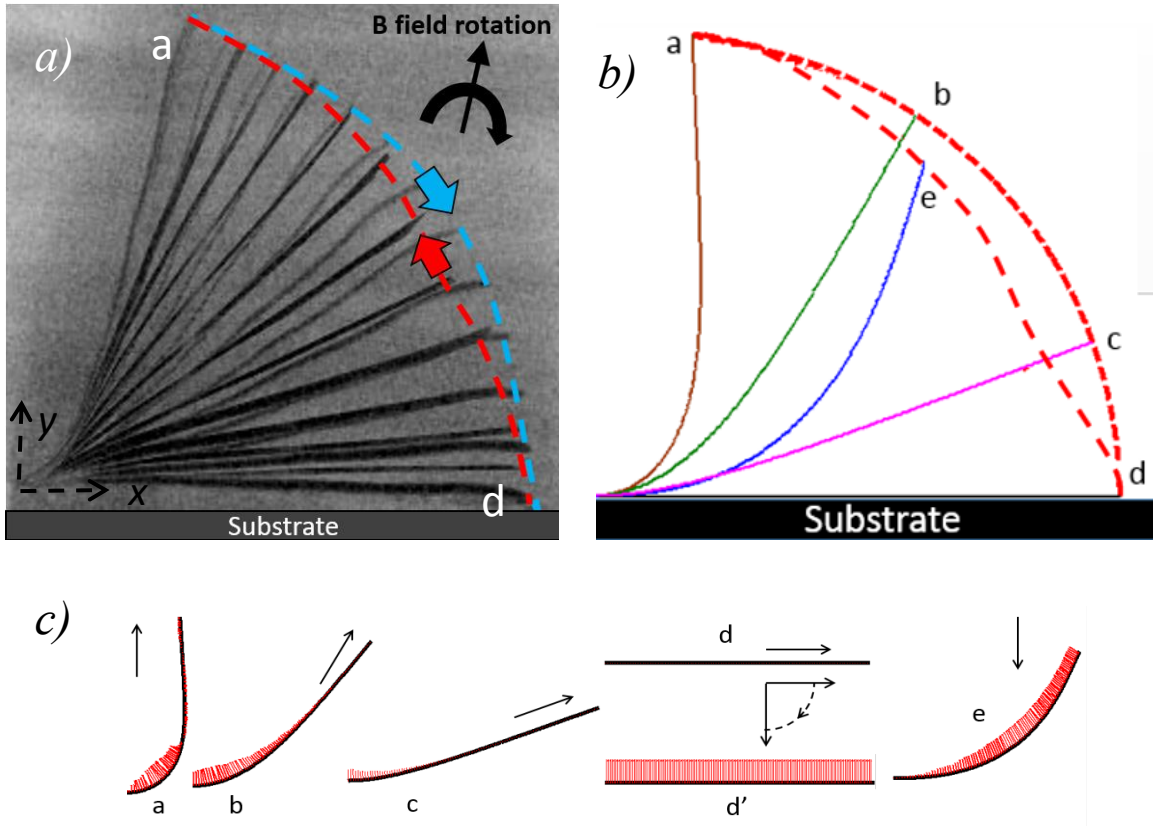

**Figure 3**: Cilium beating due to clockwise (CW) rotation of magnetic field. Experimental parameters are $L = 500\mu m$, $W = 10\mu m$, $T = 70nm$, $\omega = 0.3Hz$. (**a**) Overlay of experimental images showing cilium profiles for different orientations of magnetic field. (**b**) Computer simulation of cilium motion induced by CW rotating magnetic field. (**c**) Magnitude of local bending moment acting on the cilium at various orientations of magnetic field. The arrows indicate the field direction.

Both CW and CCW modes of cilium actuation result is two cycles of cilium beating for every rotation of the magnetic field. However, the kinematics is significantly different for these two modes. The CCW rotating magnetic field acts to increase the bending in the cilium during the forward stroke, whereas CW rotation reduces the bending in the forward stroke. These differences result in significantly different bending patterns for the CW and CCW directions of magnet rotation. Furthermore, CCW rotation leads to a significantly larger spatial asymmetry of beating cycle compared to the CW case.

It is known that spatial asymmetry in a beating pattern is essential for creating any net fluid transport at the low Reynolds number environment[31]. A direct correlation between area enclosed by the cilium tip trajectory and fluid transport has been found[33]. We, therefore, expect larger fluid transport by cilia actuated by CCW field for every oscillation cycle in these operating regimes.

*3.3. Fluid transport by ciliary array*

We characterize fluid transport by arrays of our magnetic cilia by measuring the fluid pumping in a microchannel loop. Figure 4 shows a comparison of pumping parameter $P_f$ in the microchannel loop for CW and CCW actuated cilia as a function of $Sp$. In these experiments, we use 16 rows of cilia with 25 individual cilia per row.

We find that CCW actuation results in superior pumping performance compared to CW actuation mode. This result holds for all $Sp$ tested in the experiment. The maximum pumping rate for cilia in the CCW mode is $P_f \approx 14$, which is significantly higher than the maximum pumping rate at the CW mode equal to $P_f \approx 4.5$. We relate this better pumping performance for CCW magnet rotation to a larger sweep area by the cilium tip compared to CW actuation.

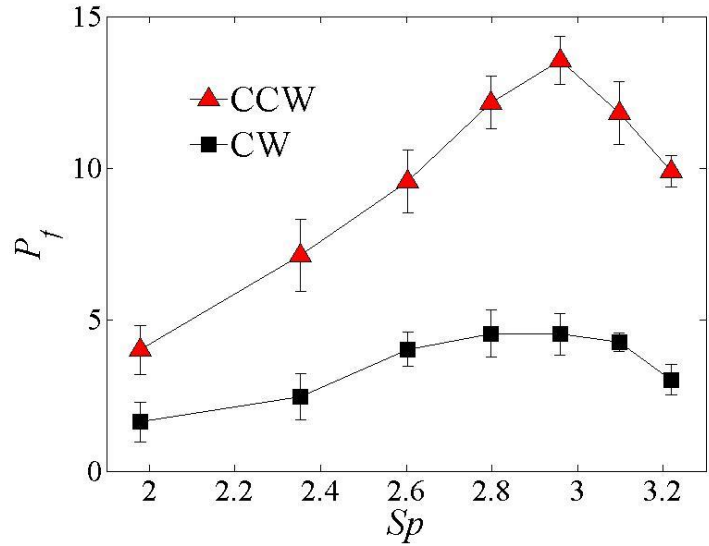

**Figure 4**: Dimensionless pumping rate $P_f$ as a function of $Sp$, for CCW and CW rotation of the magnetic field actuating a ciliary array. The array has 16 rows of cilia with 25 cilia per row. Spacing between array rows is $S = 250 \mu m$, channel height is $H = 300 \mu m$.

Furthermore, we find that the direction of flow for CCW and CW rotation of the magnet is opposite. For both CCW and CW rotation of the magnet, the cilium in forward stroke sweeps through a larger area compared to the recovery stroke. However, the direction of the forward sweep is different as indicated by the blue tip trajectories in Fig. 2a and 3a for the two rotation directions, respectively. The direction of the resulting fluid pumping coincides with the direction of the forward stroke, thus leading to different pumping directions for CCW and CW magnet rotation.

We find, for a constant magnetic number, that the pumping initially increases with increasing $Sp$ until a maximum is reached (Fig. 4). Further increase in $Sp$ results in a reduction of pumping rate per unit time. For low $Sp$ which was obtained in our experiments at lower oscillation frequencies, cilia experience weaker viscous forces. This allows the cilia to sweep a larger area increasing fluid transport[41]. The overall performance is, however, relatively low since cilia can only complete a few oscillations per unit time. As $Sp$ increases, increasing viscous forces reduce the stroke asymmetry[41] and, therefore, the fluid pumping per oscillation cycle. The faster oscillation rate, on the other hand, increases the overall pumping rate until a maximum is reached at around $Sp \approx 2.9$. Beyond this value, fluid viscosity significantly suppresses cilium beating leading to a decline of the pumping performance.

In our experiments, the maximum pumping rate obtained for CCW actuation is about $P_f \approx 14$ at 2500RPM. This corresponds to a centerline velocity equal to $\sim 1350\,\mu m/s$ and volumetric flow rate $\sim 11\,\mu l/min$ in a channel with $1mm \times 280\,\mu m$ cross-section. This flow rate exceeds the values previously reported for synthetic ciliary systems[29,33]. The self-propelling frequency $f_s = Q/S_{package}$ for our device is estimated to be $\sim 2.5\,min^{-1}$, indicating its effectiveness. Assuming a Poiseuille flow in the microchannel, we estimate a pressure drop of $\sim 1Pa$ is generated by the ciliary array in our $\sim 4cm$ channel loop.

Thus, our ciliary arrays with CCW magnetic actuation exhibit a high pumping flow rate with a low pressure drop, which is consistent with previously reported data[33]. In what follows, we will systematically examine pumping by CCW actuated cilia for different array and channel parameters.

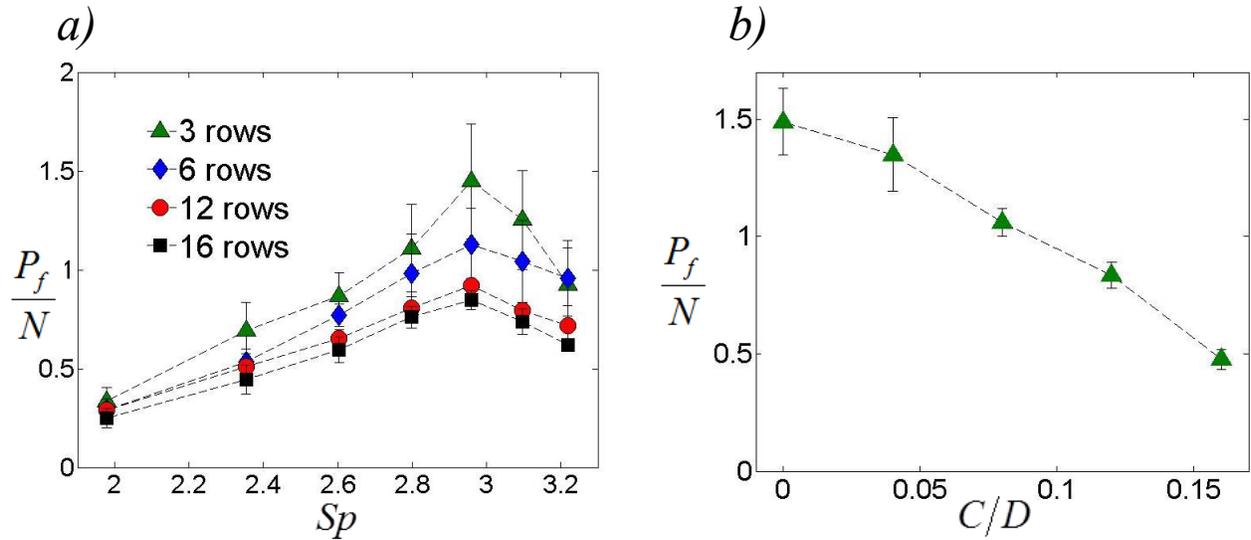

Figure 5. **a**: Dimensionless pumping rate $P_f$ produced by ciliary arrays with different number of rows as a function of $Sp$ for CCW rotation of the magnetic field actuating cilia. **b**: Pumping rate per row $P_f/N$ as a function of the normalized distance $C/D$ of the magnet from the center for an array with $N = 3$. The arrays have 25 cilia per row with spacing $S = 300\,\mu m$. Channel height is $H = 300\,\mu m$.

Figure 5 shows how the ciliary array pumping depends on the number of rows of cilia. Here, we normalize the pumping rate $P_f$ by number of rows of cilia $N$ to characterize the effectiveness per row. We vary the number of rows between 3 and 16, while keeping the separation between rows constant and equal to $300 \mu m$. We find that pumping per row $P_f/N$ decreases as the number of rows in the array increases. This result can be explained based on the difference in magnetic forces experienced by cilia at different rows. As the number of rows is increased, the distance between the magnet and cilia at different rows also changes. Cilia near the array center are closest to the magnet and experience the maximum magnetic force, whereas the cilia in the outer rows of the array experience a reduced force. In our experiments, we measure a ~ $70 Gauss$ reduction in the magnetic field strength acting on the cilia in outer rows. The reduction in magnetic field strength reduced the beating amplitude of cilia located at the outer rows in the array and, therefore, suppress the overall pumping performance of the ciliary array.

To further investigate the dependence of the cilium performance on the magnet position, we performed experiments in which we vary the magnet distance from an array. We choose a small array with $N = 3$ and displace the magnet in the $x$ direction. We define the $x$ distance between the center of the array and axis of the magnetic field as $C$. In Fig. 5b, we plot the magnitude of fluid pumping per row $P_f/N$ of the ciliary array with $N = 3$ as a function of the normalized magnet position $C/D$, where $D$ is the magnet diameter. We find a significant drop in the pumping when $C/D$ is increased from $0$ to $0.15$. For $C/D = 0.15$, the magnet is displaced by a distance of $2mm$ from the center of the array which is equivalent to the distance to the cilia near the edges for an array with $N$=16. This result suggests that the reduced pumping per row for larger arrays is indeed related to the reduced magnetic force. This decrease in performance for larger array can be mitigated by using a rotating magnet with a larger diameter to create a more uniform magnetic field across the entire array of cilia.

To understand the mutual influence of cilia in a multi-row arrays, we examine pumping by cilium arrays in which we vary the spacing $S$ (shown in Fig. 1) between consecutive rows of cilia. The cilia are actuated by CCW magnet rotation. The case where $S/L = 1$, represents a scenario in which the rows of cilia are closely packed, as can be fabricated using our method. Fig. 6 shows the pumping rate as function of $Sp$ for arrays with different inter-row spacing. We find that pumping data for the different spacing arrays differ insignificantly. In fact, the average pumping rate is within the experimental error for all $S/L$. This implies that the cilium performance is not significantly affected by neighboring cilium rows and the hydrodynamic interaction between them is relatively weak.

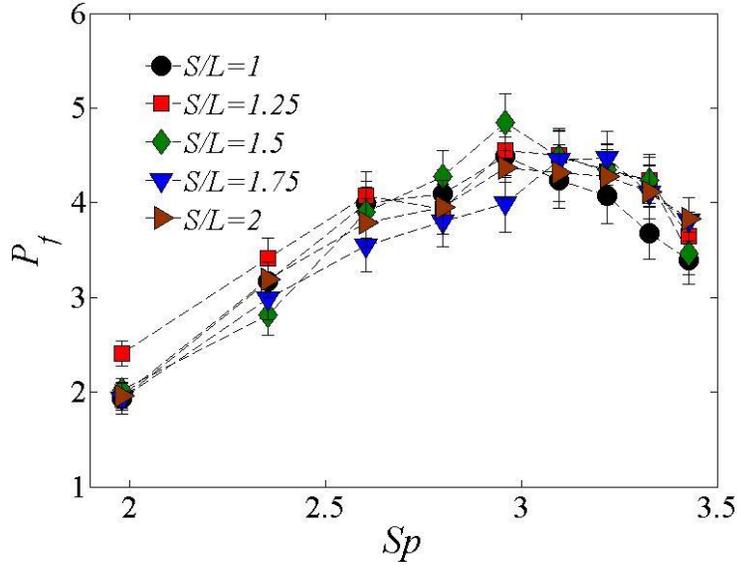

**Figure 6**. Dimensionless pumping rate $P_f$ as a function of $Sp$ for ciliary arrays with different spacing between consecutive rows. All arrays have four rows of cilia with 25 cilia per row. Channel height is $H = 300 \mu m$.

In our final set of experiments, we probe how pumping by cilia depends on the channel height $H$. Figure 7 shows the effect of varying channel height on the cilium array pumping. In these experiments, we examine cilia in CCW mode with three different $Sp$. The cilia are arranged in 4 rows with 25 cilia per row. The channel height is varied between $220 \mu m$ to $750 \mu m$. We find a significant increase in pumping rate from $P_f \approx 2.8$ to $P_f \approx 4$, as $H/L$ is increased from 1.2 to 1.8. Further increase in channel height beyond $H/L = 1.8$ results in a little change in pumping performance.

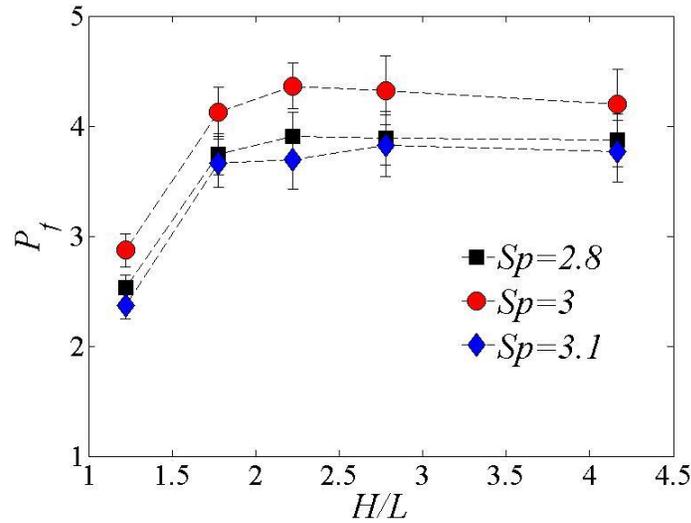

**Figure 7**. Dimensionless pumping rate $P_f$ as a function of normalized channel height $H/L$ for selected values of $Sp$. A ciliary array with four rows of cilia, and 25 cilia per row is used. Spacing between rows of cilia is $S = 250 \mu m$.

It is predicted that the flow above the cilia in a closed loop channel with $H \gg L$ resembles the Couette flow, with no-slip at the top boundary and bottom boundary being tangentially moved by the cilia[33]. In such cases, the flowrate scales linearly with channel height and the centerline velocity is independent of channel height. Our experiments confirm this result, as demonstrated by the negligible change in centerline velocity for $H/L \geq 1.8$. However, when channel height is comparable to cilia length, the centerline velocity decreases as shown for $H/L = 1.2$. This indicates that the motion of the cilia is affected by the channel top wall.

## 4. Conclusions

We demonstrate fluid pumping by an array of beating magnetic cilia. The cilia are made of soft magnetic thin films, which are actuated by a rotating permanent magnet. Kinematics of a single cilium is examined by imaging its motion. The cilium follows different bending pattern defined by the direction of magnet rotation. The CCW rotation of the magnet produces larger spatial asymmetry in the beating pattern compared to the CW rotation of the magnet. This, in turn, results in faster pumping by cilia actuated using CCW rotating magnet. The pumping is shown to depend on a dimensionless sperm number with the fastest pumping rate occurring at $Sp \approx 2.9$ The pumping rate increases with increasing the number of ciliary rows in the array, but nearly independent from the spacing between the rows. The maximum centerline velocity generated by our magnetic cilia is $\sim 1350\,\mu m/s$, which corresponds to a volumetric flowrate of $\sim 11\,\mu l/min$. The self-pumping frequency, a metric used to assess the effectiveness of the pump based on its size, is estimated to be $\sim 2.5\,min^{-1}$. This is the highest reported value in similar ciliary microfluidic pumping systems. The flowrates produced by our artificial cilia are comparable to that of natural cilia[9,14,15].

Artificial magnetic cilia such as developed in this study are promising for active control of fluid flow in microfluidic devices. Our experimental results provide guidelines for designing biomimetic magnetic cilia for various lab-on-a-chip applications. The magnetic actuation has the particular advantage of being able to actuate remotely and not interfere with biological samples. Such cilia incorporated inside stents and tubes can be used for precise fluid metering. Furthermore, combining magnetic cilia with different orientations can be used for fluid mixing and probing the behavior of microorganisms in a variety of flow conditions.

**Acknowledgments**

We thank the USDA NIFA (grant #11317911) and the National Science Foundation (CBET-1510884) for financial support and the staff of Georgia Tech IEN for assistance with clean room fabrication.